\documentclass[a4paper,10pt]{article}
\usepackage{graphicx,hyperref}
\usepackage{amsmath,amsfonts,amssymb}
\usepackage{fullpage}
\usepackage{authblk}
\usepackage{setspace}
\usepackage{subcaption}
\usepackage{cite}
\usepackage{makecell}

\usepackage{color}

\title{\bf{Evolutionary dynamics of the cryptocurrency market}}

\author[a]{Abeer ElBahrawy}
\author[a]{Laura Alessandretti}
\author[b]{Anne Kandler}
\author[c]{Romualdo Pastor-Satorras}
\author[a,d,*]{Andrea Baronchelli}

\affil[a]{{\small Department of Mathematics - City, University of London - Northampton Square, London EC1V 0HB, UK}}

\affil[b]{{\small Max Planck Institute for Evolutionary Anthropology, Department of Human Behavior, Ecology and Culture, Leipzig, Germany}}

\affil[c]{{\small Departament de F\'\i sica, Universitat Polit\`ecnica de
  Catalunya, Campus Nord B4, 08034 Barcelona, Spain}}
  
\affil[d]{{\small UCL Centre for Blockchain Technologies, University College London, UK} 

{\small $^*$Corresponding author:  Andrea.Baronchelli.1@city.ac.uk}}

\date{}
\begin{document}
\maketitle

\begin{abstract}
The cryptocurrency market surpassed the barrier of $\$100$ billion market capitalization in June 2017, after months of steady growth. Despite its increasing relevance in the financial world, however, a comprehensive analysis of the whole system is still lacking, as most studies have focused exclusively on the behaviour of one (Bitcoin) or few cryptocurrencies.  Here, we consider the history of the entire market and analyse the behaviour of $1,469$ cryptocurrencies introduced between April 2013 and June 2017. We reveal that, while new cryptocurrencies appear and disappear continuously and their market capitalization is increasing (super-) exponentially, several statistical properties of the market have been stable for years. These include the number of active cryptocurrencies, the market share distribution and the turnover of cryptocurrencies. Adopting an ecological perspective, we show that the so-called neutral model of evolution is able to reproduce a number of key empirical observations, despite its simplicity and the assumption of no selective advantage of one cryptocurrency over another.  Our results shed light on the properties of the cryptocurrency market and establish a first formal link between ecological modelling and the study of this growing system.  We anticipate they will spark further research in this direction.

\end{abstract}


\section{Introduction}
Bitcoin is a digital asset designed to work as a medium of exchange \cite{nakamoto2008bitcoin, ali2014economics}. Users can send and receive native tokens - the ``bitcoins'' - while collectively validating the transactions in a decentralized and transparent way. The underlying technology is based on a public ledger - or blockchain - shared between participants and a reward mechanism in terms of bitcoins as an incentive for users to run the transaction network. It relies on cryptography to secure the transactions and to control the creation of additional units of the currency, hence the name of ``cryptocurrency'' \cite{pagliery2014bitcoin,vigna2015age}.

After Bitcoin appeared in $2009$, approximately $1,500$ other
cryptocurrencies have been introduced, around $600$ of which are
actively traded today. All cryptocurrencies share the underlying
blockchain technology and reward mechanism, but they typically live on
isolated transaction networks. Many of them are basically clones of
Bitcoin, although with different parameters such as different supplies,
transaction validation times, etc. Others have emerged from more
significant innovations of the underlying blockchain technology
\cite{camb2017} (see \ref{sec:PoWPoS}).

Cryptocurrencies are nowadays used both as media of exchange for daily
payments, the primary reason for which Bitcoin was introduced, and for
speculation \cite{ceruleo2014bitcoin,rogojanu2014issue}.  Other uses
include payment rail for non-expensive cross borders money transfer and
various non-monetary uses such as time-stamping \cite{ali2014economics}.
The self-organization of different usages both within a single
cryptocurrency and as an element of differentiation between
cryptocurrencies makes the market of cryptocurrencies unique, and their
price extremely volatile
\cite{yermack2013bitcoin,kristoufek2015main,RePEc:cto:journl:v:35:y:2015:i:2:p:383-402}.

Between $2.9$ and $5.8$ millions of private as well as institutional
users actively exchange tokens and run the various transaction networks
\cite{camb2017}.  In May $2017$, the market capitalization of active
cryptocurrencies surpassed $\$91$ billion \cite{coincap}. Bitcoin
currently dominates the market but its leading position is challenged
both by technical concerns
\cite{reid2013analysis,bamert2013have,decker2013information,gervais2014bitcoin,BitcoinBlockDebate}
and by the technological improvements of other cryptocurrencies
\cite{wang2017buzz}.

Despite the theoretical and economic interest of the cryptocurrency
market \cite{ali2014economics,vigna2015age,casey2015bitcoin,trimborn2016crix}, however, a
comprehensive analysis of its dynamics is still lacking. Existing
studies have focused either on Bitcoin, analysing for example the
transaction network
\cite{ron2013quantitative,kondor2014rich,tasca2016evolution,kondor2014inferring,lischke2016analyzing}
or the behaviour and destiny of its price
\cite{iwamura2014bitcoin,kristoufek2013bitcoin,kristoufek2015main,cusumano2014bitcoin,garcia2015social,garcia2014digital,ciaian2016economics},
or on a restricted group of cryptocurrencies (typically $5$ or $10$) of
particular interest \cite{gandal2016can,camb2017,wang2017buzz,elendner2016cross}. But even
in this case there is disagreement as to whether Bitcoin dominant
position may be in peril \cite{camb2017} or its future dominance as
leading cryptocurrency is out of discussion \cite{gandal2016can}. 

Here we present a first complete analysis of the cryptocurrency market,
considering its evolution between April 2013 and June 2017. We focus on the
market shares of the different cryptocurrencies (see \ref{Methods}) and find that Bitcoin has been steadily
losing ground to the advantage of the immediate runners-up. We then show
that several statistical properties of the system have been stable for
the past few years, including the number of active cryptocurrencies, the
market share distribution, the stability of the ranking, and the birth
and death rate of new cryptocurrencies. We adopt an ``ecological''
perspective on the system of cryptocurrencies and notice that several
observed distributions are well described by the so-called ``neutral
model'' of evolution \cite{kimura1983neutral,Bire82}, which also captures
the decrease of Bitcoin market share. We believe that our findings
represent a first step towards a better understanding and modelling of
the cryptocurrency market.

\begin{figure}[!h]
\centering\includegraphics[width=2.5in]{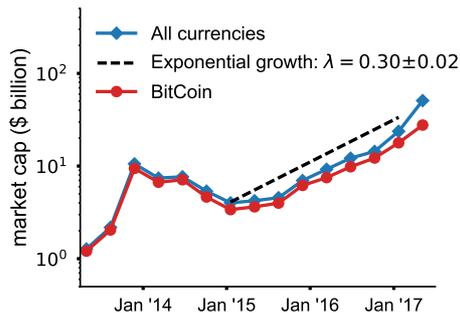}
\caption{\textbf{Evolution of the market capitalization.} Evolution of the market capitalization over time (starting from April 2013), for all cryptocurrencies (blue line,diamonds) and for Bitcoin (red line, dots). The dashed line is an exponential curve $f(t) \sim e^{\lambda t }$, with $\lambda = 0.3$, shown as a guide for the eye. Data is averaged over a 15-week window.}
\label{fig:MCEvol}
\end{figure}

\section{Results}
\subsection{Market Description \label{sec:ResMarketDes}}

Our analysis focuses on the market share of the different
cryptocurrencies and is based on the whole history of the cryptocurrency
market between April 28, 2013 and May 13, 2017.  Our dataset
includes $1,469$ cryptocurrencies, of which around $600$ were active by that time (see \ref{Methods}).

The total market capitalization $C$ of cryptocurrencies has been
increasing since late 2015 after a period of relative tranquillity
(Fig.~\ref{fig:MCEvol}). As of May 2017, the market capitalization is
more than $4$ times its value compared to May 2016 and it exhibits an
exponential growth $C \sim \exp(\lambda t)$ with coefficient
$\lambda = 0.30 \pm 0.02 $, where $t$ is measured in units of 15 weeks.

\subsection{Decreasing Bitcoin Market Share}
\label{sec:bitcoin}
Bitcoin was introduced in 2009 and followed by a second cryptocurrency
(Namecoin, see \ref{sec:topcryp}) only in April 18,
2011. This first-mover advantage makes Bitcoin the most famous and
dominant cryptocurrency to date. However, recent studies analysing the
market shares of Bitcoin and other cryptocurrencies reached contrasting
conclusions on its current state. While Gandal and Halaburdain in their
2016 study concluded that ``Bitcoin seems to have emerged - at least in
this stage - as the clear winner'' \cite{gandal2014competition}, the
2017 report by Hileman and Rauchs noted that ``Bitcoin has ceded
significant market cap share to other cryptocurrencies''
\cite{camb2017}.

\begin{figure}[!h]
\centering\includegraphics[width=5in]{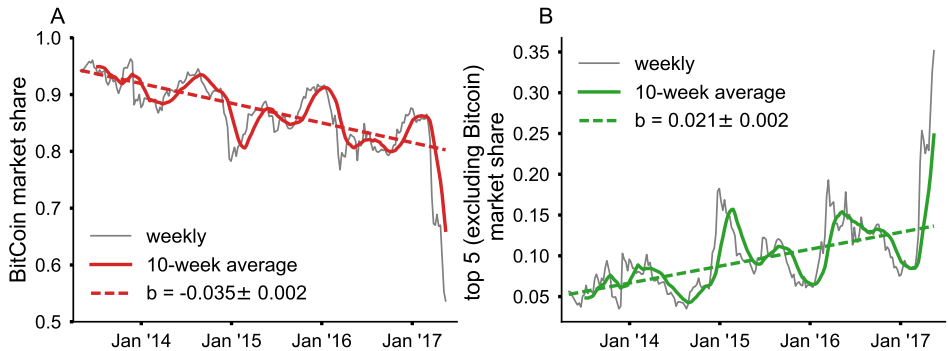}
\caption{\textbf{Evolution of the market share
      of top-ranking cryptocurrencies.} \textbf{(A)} The market share of
    Bitcoin across time sampled weekly (gray line) and averaged over a
    rolling window of $10$ weeks (red line). The dashed line is a linear
    fit with angular coefficient $b=-0.035 \pm 0.002$ (the rate of
    change in 1 year) and coefficient of determination $R^2=0.63$. The
    Spearman correlation coefficient is $\rho = -0.8$, revealing a
    significant negative correlation at significance level of
    $1\%$. \textbf{(B)} Total market share of the top 5 cryptocurrencies
    excluding Bitcoin sampled weekly (gray line) and averaged over a
    rolling window of $10$ weeks (green line). The dashed line is a
    linear fit with angular coefficient $b=0.021 \pm 0.002$ (the rate of
    change in 1 year) and coefficient of determination $R^2=0.45$. The
    Spearman correlation coefficient is $\rho = 0.67$, revealing a
    significant positive correlation at significance level of $1\%$.}
\label{fig:BitcoinDec}
\end{figure}

To clarify the situation, we consider the whole evolution of the Bitcoin
market share over the past $4$ years. Fig. \ref{fig:BitcoinDec}A shows
that Bitcoin market share has been steadily decreasing for the past
years, beyond oscillations that might mask this trend to short-term
investigations. The decrease is well described by a linear fit
$f(t) = a + bt$ with angular coefficient $b = -0.035 \pm 0.002$
representing the change in market share over $t=1$ year. Neglecting the impact of non-linear effects and potential changes in the competition environment, the model indicates that Bitcoin market share can fluctuate around $50\%$ by 2025. Conversely, Fig.
\ref{fig:BitcoinDec}B shows that the top 5 runners-up (see \ref{sec:topcryp}) have
gained significant market shares and now account for more than $20\%$ of the market. 

\subsection{Stability of the Cryptocurrency Market}
\label{sec:stability}

In order to characterize the cryptocurrencies dynamics
better, we now focus on the statistical properties of the market. We
find that while the relative evolution of Bitcoin and rival
cryptocurrencies is tumultuous, many statistical properties of the
market are stable.

\begin{figure}[!h]
\centering\includegraphics[width=5in]{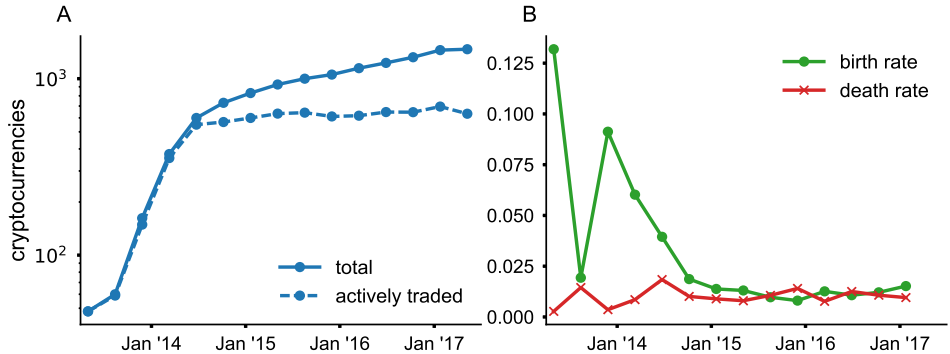}
\caption{\textbf{Evolution of the number of
      cryptocurrencies.} \textbf{(A)} The number of cryptocurrencies
    that ever entered the market (filled line) since April 2013, and the
    number of actively traded cryptocurrencies (dashed line). \textbf{(B)} The
    birth and death rate computed across time. The birth (resp. death)
    rate is measured as the fraction of cryptocurrencies entering
    (resp. leaving) the market on a given week over the number of living
    cryptocurrencies at that point. Data is averaged over a 15 weeks
    window.}
\label{fig:NumCyrptEvol}
\end{figure}

Fig.~\ref{fig:NumCyrptEvol}A
shows the evolution of the number of active cryptocurrencies across
time, averaged over a $15$-week window. The number of actively traded
cryptocurrencies is stable due to similar birth and death rates since
the end of 2014 (Fig.~\ref{fig:NumCyrptEvol}B). The average monthly
birth and death rates since 2014 are 1.16\% and 1.04\%, respectively,
corresponding to approximately $7$ cryptocurrencies appearing every week
while the same number is abandoned.

\begin{figure}[!h]
\centering\includegraphics[width=5in]{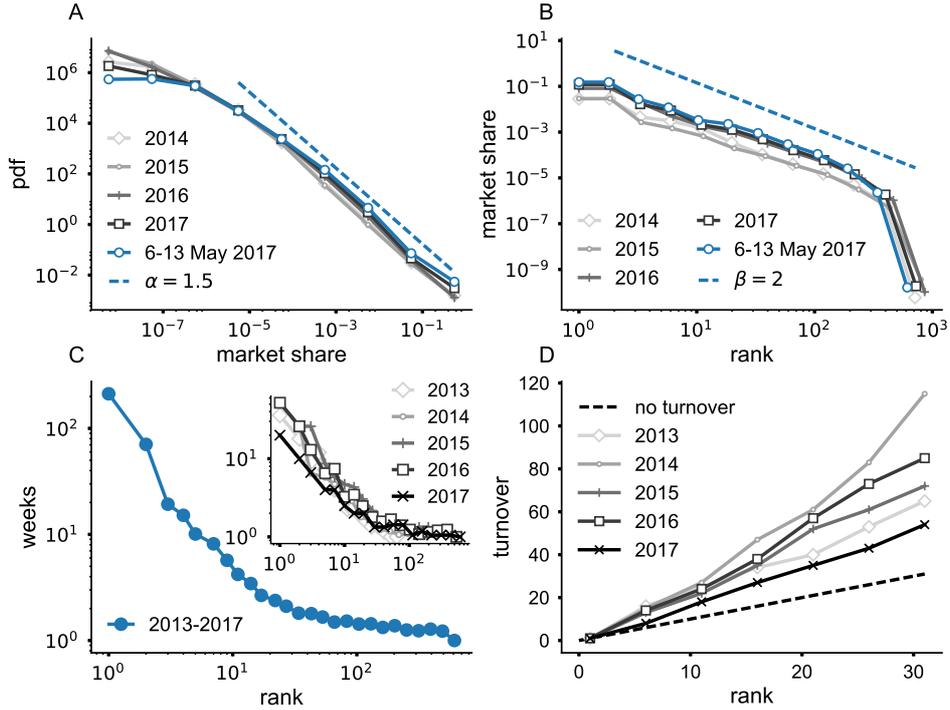}
\caption{\textbf{Stable properties of
      the cryptocurrency market.} \textbf{(A)} Distribution of market share computed
    aggregating across a given year (gray filled lines), and over the
    week 6-13 May 2017 (blue thick line). The dashed line is a power law
    $P(x) \sim x^{-\alpha}$ curve with exponent $\alpha =
    1.5$. \textbf{(B)} Frequency-rank distribution of cryptocurrencies,
    computed aggregating across a given year (gray filled lines), and
    over the week 6-13 May 2017 (blue thick line). The dashed line is a
    power law curve $P(r) \sim r^{-\beta}$ with exponent $\beta = 2$.  \textbf{(C)} Average amount of time (in weeks) a
    cryptocurrency occupies a given rank computed averaging across all
    years (blue line), and across given years (gray lines, inset). \textbf{(D)}
    Turnover of the ranking distribution, defined as the total number of
    cryptocurrencies ever occupying rank higher than a given rank. The
    measure is computed averaging across given years (gray filled
    lines). The 2013 and 2017 curves must be taken purely as an indication as they
    are computed on less than $12$ months (approximately $8$ and $4$ months, respectively). The dashed
    line has angular coefficient $1$, and corresponds to the case in which the ranking of
    cryptocurrencies is fixed (i.e., the variable turnover captures only the initial size of the toplist).}
\label{fig:DistDataYear}
\end{figure}

Interestingly, the market share distribution remains stable across
time. Fig. \ref{fig:DistDataYear}A shows that curves obtained by
considering different periods of time are indistinguishable. This is
remarkable because the reported curves are obtained by considering data
from different years as well as data aggregated on different time spans
- from one week to the entire $\sim4$ years of data. The obtained
distribution exhibits a broad tail well described by a power law
$P(x) \sim x^{-\alpha}$ with exponent $\alpha = 1.58 \pm 0.12$ (Fig.\ref{fig:DistDataYear}A), where the fit coefficient is computed using the method detailed in \cite{clauset2009power}. The expected relationship between the probability distribution and the frequency rank distribution predicts the latter is a power-law function $P(r) \sim r^{-\beta}$ with exponent $\beta = 1/(\alpha -1)$ \cite{adamic2002zipf}, yielding in our case $\beta =1.72$ Fig.\ref{fig:DistDataYear}B. The empirical fit coefficient $\beta = 1.93\pm 0.23$ is consistent with this prediction. (Fig.
\ref{fig:DistDataYear}B). This was also verified for each year individually (see \ref{ver_exp}).

We further investigate the stability of the market by measuring the
average rank occupation time (Fig.~\ref{fig:DistDataYear}C),
defined as the amount of time a cryptocurrency typically spends in a
given rank before changing it. We find that the time spent in a top-rank
position decays fast with the rank, while for low-rank positions such
time approaches 1 week. Again, this behaviour is stable across years
(Fig.~\ref{fig:DistDataYear}C - inset). We also consider the
turnover profile defined as the total number of cryptocurrencies ever occupying rank higher than a given rank in period $t$ (see \cite{bentley2007regular} for a similar definition). Fig.~\ref{fig:DistDataYear}D shows that also this quantity is substantially stable across time.

The first rank has been always occupied and continues to be occupied by
Bitcoin, while the subsequent $5$ ranks (i.e., ranks $2$ to $6$) have
been populated by a total of $33$ cryptocurrencies with an average
life time of $12.6$ weeks. These values change rapidly when we consider
the next set of ranks from $7$ to $12$ to reach $70$ cryptocurrencies
and an average life time of $3.6$ weeks. At higher ranks, the mobility
increases and cryptocurrencies continuously change position.

\subsection{A Simple Model for the Cryptocurrency Ecology}
\label{Neutral Model}

In order to account for the empirical properties of the dynamics of cryptocurrencies we have discussed above, we adopt the view of a ``cryptocurrency ecology'' and consider the neutral model of evolution, a prototypical model in population-genetics and ecology \cite{kimura1983neutral,Bire82}.

The Wright-Fisher model of neutral evolution describes a fixed size population of $N$ individuals where each individual belongs to one of $m$ species. At each generation, the $N$ individuals are replaced by $N$ new individuals. Each new individual belongs to a species copied at random from the previous generation, with probability $1-\mu$, or to a species not previously seen, with probability $\mu$, where $\mu$ is a mutation parameter that does not change over time \cite{ewens2012mathematical}. Despite its simplicity, the neutral model is able to reproduce the static patterns of the competition dynamics of many systems including ecological \cite{Mcgill2006empirical} and genetics \cite{kimura1968evolutionary} systems, cultural change
\cite{neiman1995stylistic}, English words usage
\cite{Ruck2017neutral} and technology patents citations
\cite{Bentley2004random}. 

In our mapping of the ecological model to the cryptocurrency market,
each individual corresponds to a certain amount of dollars, while species correspond to different
cryptocurrencies (see \ref{sec:sim}). The copying mechanism represents trading, with $\mu$
denoting the probability that a new cryptocurrency is
introduced. Our choice of $\mu$ is informed by the data to yield a
number of new cryptocurrencies per unit time corresponding to the
empirical observation. We thus fix $\mu = \frac{7}{N}$, where $N$ is the
population size in the model. Thus, one model generation corresponds to
$1$ week of observations, the choice of $\mu$ guaranteeing an average of
$7$ new cryptocurrencies entering the system every week, as empirically
observed. Finally, in contrast to most neutral models, we assume that a
new species does not enter the system with a single individual but with
a size proportional to the empirical average market share of new
cryptocurrency (see \ref{sec:sim}).

\begin{figure}[!h]
\centering\includegraphics[width=5in]{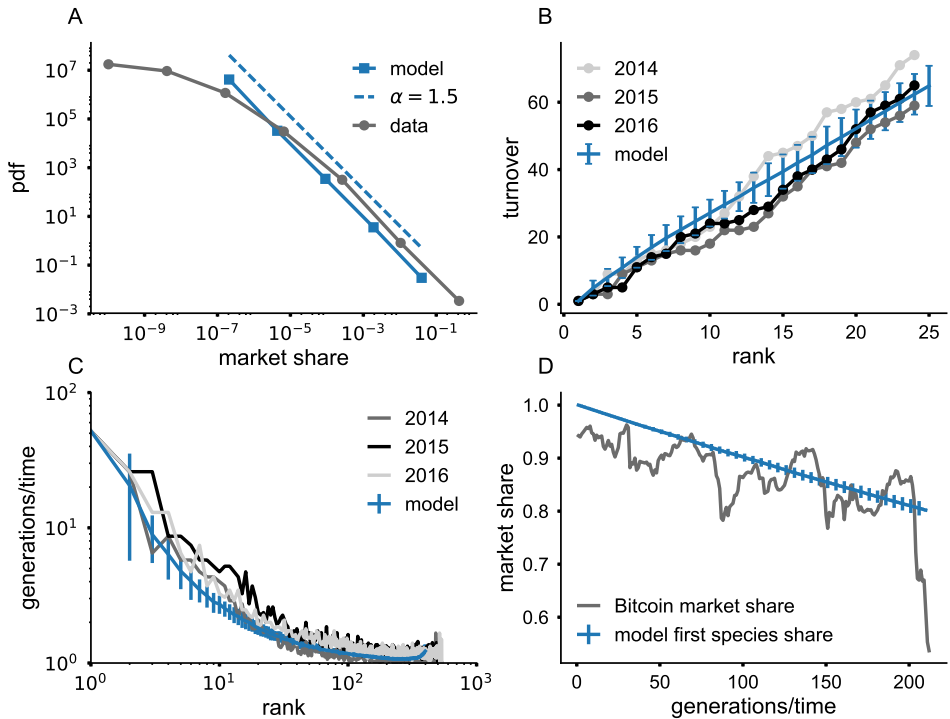}
\caption{\textbf{Neutral model for evolution
        and empirical observations.} \textbf{(A)} Distribution of
      cryptocurrencies market shares aggregated over all years (gray
      line, dots) and the equilibrium distribution resulting from
      numerical simulations (blue line, squares) aggregated over $210$
      generations. The dashed line is the power law curve
      $P(x) \sim x^{-\alpha}$ predicted analytically with exponent
      $\alpha = 1.5$ \cite{O2017novelty}. \textbf{(B)} Turnover of the
      ranking distribution computed considering $52$ generations for the
      cryptocurrencies data (gray lines, dots) and for numerical
      simulations (blue line), \textbf{(C)} Average number of
      generations a cryptocurrency (gray lines) and a species in the
      neutral model (blue line) occupies a given rank. Averages are
      computed across $52$ generations. \textbf{(D)} Evolution of the
      market share of Bitcoin (gray line) and the expected market share of the
      first species in numerical simulations (blue line). All
      simulations are run for $N = 10^{5}$ and $\mu = 7/N$ starting from
      1 species in the initial state.  The size of entering species $m$,
      whose average $m=15$ is informed by the data, is taken at random
      in the interval $m=[10,20]$. Error bars are standard deviations,
      computed across 100 simulations. For
      panels (B) and (C) measures start at generation $g_1 =105$ (see \ref{sec:sim} for variations of this parameter).}
\label{fig:MSFirstSim}
\end{figure}

The neutral model translates in the simplest way three main assumptions
\cite{alonso2006merits}: (i) interactions between cryptocurrencies are
equivalent on an individual per capita basis (i.e., per US dollar); (ii)
the process is stochastic; and (iii) it is a sampling theory, where the
new generation is the basis to build the following one. In other words,
the neutral model assumes that all species/cryptocurrencies are
equivalent and that all individuals/US dollars are equivalent.

Testing the consistency between observed patterns of the cryptocurrency market and theoretical expectations of neutral theory revealed that neutrality captures well at least four features of the cryptocurrency ecology, namely:

\begin{enumerate}
\item The exponent of the market share distribution (Fig
  \ref{fig:MSFirstSim}A);
\item The linear behavior of the turnover profile of the dominant
  cryptocurrencies (Fig \ref{fig:MSFirstSim}B);
 \item The average occupancy time of any given rank (Fig \ref{fig:MSFirstSim}C);
\item The linear decrease of the dominant cryptocurrency (Fig
  \ref{fig:MSFirstSim}D).
\end{enumerate}

The neutral model generates in fact an aggregated species distribution (i.e., obtained when all generations up to the $i^{th}$ are combined together and analysed as a single population of size $N*i$  \cite{hahn2003drift,Bentley2004random}) that, at equilibrium, can be described by a power law distribution $P(x) \sim x^{-\alpha}$ with $\alpha = 1.5$ \cite{O2017novelty}, in agreement with the empirical value $\alpha = 1.58 \pm 0.12$ obtained by the fitting procedure described in \cite{clauset2009power}. Fig. \ref{fig:MSFirstSim}A shows the agreement between simulations and data (same behaviour of the long tail), where simulations results are aggregated over $i=210$ generations, corresponding to $4$ years of empirical observations under our choice of $\mu$. The existence of a power law phase with exponent $1.5$ in the model is independent of $\mu$ (see \ref{sec:sim}) \cite{O2017novelty}. 

Furthermore, when we account for the fact that Bitcoin was originally the only cryptocurrency by setting 1 species in the initial state, the model captures also the remaining properties. In Fig. \ref{fig:MSFirstSim}B and \ref{fig:MSFirstSim}C, we compare the turnover profile and the ranking occupation times with the corresponding simulation results. We compute these quantities over a period of $52$ generations, corresponding to one year of observations. The curves reported in Figs. \ref{fig:MSFirstSim}B and \ref{fig:MSFirstSim}C correspond to measures performed between generation $g_1=105$ and $g_2=156$, corresponding to year $3$ ($2015$) in the data. Crucially, however, both measures are stable in time, i.e. they do not depend on the choice of $g_1$ (but for an initial period of high rank variability for the very first generations, see \ref{sec:sim}). It is worth noting that the linearity of the turnover profile in Fig.~\ref{fig:MSFirstSim}B corresponds to a similar behaviour observed in \cite{bentley2007regular} when the measure is performed between two consecutive generations. Fig.~\ref{fig:MSFirstSim}D shows the observed linear decrease of the leading cryptocurrency market share (Fig.~\ref{fig:MSFirstSim}C), indicating that newborn cryptocurrencies mostly damage the dominating one.



\section{Discussion and Outlook}
\label{Discussion}

In this paper we have investigated the whole cryptocurrency market between
April 2013 and June 2017. We have shown that the total market capitalization has
entered a phase of exponential growth one year ago, while the market
share of Bitcoin has been steadily decreasing. We have identified
several observables that have been stable since the beginning of our
time series, including the number of active cryptocurrencies, the
market-share distribution and the rank turnover. By adopting an
ecological perspective, we have pointed out that the neutral model of
evolution captures several of the observed properties of the market.

The model is simple and does not capture the full complexity of the
cryptocurrency ecology. However, the good match with at least part of
the picture emerging from the data does suggest that some of the
long-term properties of the cryptocurrency market can be accounted for
based on simple hypotheses. In particular, since the model assumes no
selective advantage of one cryptocurrency over the other, the fit with
the data shows that there is no detectable population-level consensus on
what is the ``best'' currency or that different currencies are
advantageous for different uses. Furthermore, the matching between 
the neutral model and the data implies that the observed patterns
of the cryptocurrency market are compatible with a scenario where
technological advancements have not been key so far (see \ref{sec:PoWPoS}) and where users and/or investors
allocate each packet of money independently.
Future work will need to consider the
role of an expanding overall market capitalization and, more
importantly, try to include the information about single transactions,
where available, in the modelling picture.

In the immediate and mid-term future, legislative, technical and social
advancements will most likely impact the cryptocurrency market
seriously and our approach, together with recent results in computational social science dealing with the quantification of financial trading and bubble formation \cite{preis2011switching,botta2015quantifying,sornette2001significance,johnson2013abrupt}, could help make sense of the market evolution. In April $2017$, for example, Japan started treating
Bitcoin as a legal form of payment driving a sudden increase in the
Bitcoin price in US dollars \cite{japan} while in February $2017$ a change of
regulation in China resulted to a
\$$100$ price drop \cite{china}. Similarly, the exponential increase in
the market capitalization (Fig. \ref{fig:MCEvol}) will likely attract further speculative attention towards this
market while at the same time increasing the usability of cryptocurrencies
as a payment method. While the use of cryptocurrencies as speculative
assets should promote diversification \cite{gandal2016can}, their
adoption as payment method (i.e., the conventional use of a shared
medium of payment) should promote a winner-take-all regime
\cite{baronchelli2017emergence,lewis2008convention}. How the
self-organized use of cryptocurrencies will deal with this tension is an
interesting question do be addressed in future studies.

\section{Material and methods}
\label{Methods}
\subsection{Data} 
Cryptocurrency data was extracted from the website Coin Market Cap
\cite{coincap}, collecting weekly data from $157$ exchange markets
platforms starting from April 28, 2013 up to May 13, 2017. For all living cryptocurrencies,
the website provides the market capitalization, the price in
U.S. dollars and the volume of trading in the preceding 24 hours. Data
on trading volume was collected starting from December 29, 2013.

The website lists cryptocurrencies traded on public exchange markets
that are older than 30 days and for which an API as well as a public URL
showing the total mined supply are available.  Information on the market
capitalization of cryptocurrencies that are not traded in the 6 hours
preceding the weekly release of data is not included on the
website. Cryptocurrencies inactive for $7$ days are not included in the
list released. These measures imply that some cryptocurrencies can
disappear from the list to reappear later on.

\subsection{Analysis} 

The following quantities characterize individual cryptocurrencies: The
\emph{circulating supply} is the number of coins available to users. The
\emph{price} is the exchange rate, determined by supply and demand
dynamics. The \emph{market capitalization} is the product of the
circulating supply and the price. The \emph{market share} is the market
capitalization of a currency normalized by the total market
capitalization.

Most of our analyses consider the market capitalization and market share
of cryptocurrencies. These quantities neglect the destroyed or dormant
coins, accounting for example to 51\% of mined Bitcoins based on data
from the period July 18, 2010 to May 13, 2012
\cite{ron2013quantitative}.

\section*{Data accessibility}
Dataset used in this study is public and can be found in Coin Market Cap
\cite{coincap}.

\section*{Competing interests}
The authors declare no competing financial interests.

\section*{Authors' contributions}
Study conception: AB;  Study design: AE, LA, AK, RPS, AB; Acquisition of data and pre-processing: AE; Analysis and interpretation of data: AE, LA, AK, RPS, AB; Drafting of manuscript: AE, LA, AK, RPS, AB;

\section*{Funding}

R.P.-S. acknowledges financial support from the Spanish MINECO, under
projects FIS2013-47282-C2- 2 and FIS2016-76830-C2-1-P, and additional
financial support from ICREA Academia, funded by the Generalitat de
Catalunya.

\bibliographystyle{unsrt}

\appendix
\setcounter{figure}{0}

\renewcommand\thefigure{\thesection\arabic{figure}}   
\section{Appendix}

\subsection{some relevant cryptocurrencies}
\label{sec:topcryp}

Table \ref{tab:table} provides information on some relevant cryptocurrencies, either occupying high-rank positions or early introduced in the market. Data was collected in May 2017, see below for details on the Technology column. 
\renewcommand{\arraystretch}{1.5}
\begin{table*}[!h]
\centering
\caption{\label{tab:table}  \textbf{Details on the top runner cryptocurrencies in the market.} The table is generated using data collected on May 28, 2013}
\begin{tabular}{ l l l l l l} 
 \hline
\textbf{Name} & \textbf{Year} & \textbf{Technology} & \textbf{Market Cap (\$)} & \textbf{Rank} & \textbf{Additional Info} \\
\hline
\textbf{Bitcoin} & 2009 & Proof-of-work & 35B & 1 &  \\
\textbf{Ethereum} & 2015 & Proof-of-work & 15B  & 2  & Smart contracts \\
\textbf{Ripple} & 2013 &  \makecell[tl]{Distributed open source \\consensus ledger} & 8B  & 3 & \makecell[tl]{ Widely adopted by \\ companies and banks.} \\
\textbf{NEM} & 2015 & Proof-of-importance & 1B  & 4 &  \\
\textbf{Ethereum Classic} & 2015 & Proof-of-work & 1B  & 5 &  \makecell[tl]{DAO Hard-fork} \\
\textbf{Litecoin} & 2011 & Proof-of-work & 1B & 6 &  \\
\textbf{Dash} & 2014 & Proof-of-work & 809M & 7  & \makecell[tl]{Gained market since \\ early $2017$. Privacy focused.} \\
\textbf{Monero} & 2014 & Proof-of-work & 535M  & 8  & \makecell[tl]{Gained momentum in late \\ $2016$. Privacy focused} \\
\textbf{NameCoin} & 2015 & Proof-of-work & 21M  & 58 &  \\
\end{tabular}
\end{table*}

\subsection{Simulations}
\label{sec:sim}

\begin{figure}[!h]
  \begin{center}
    \includegraphics[width=2.5in]{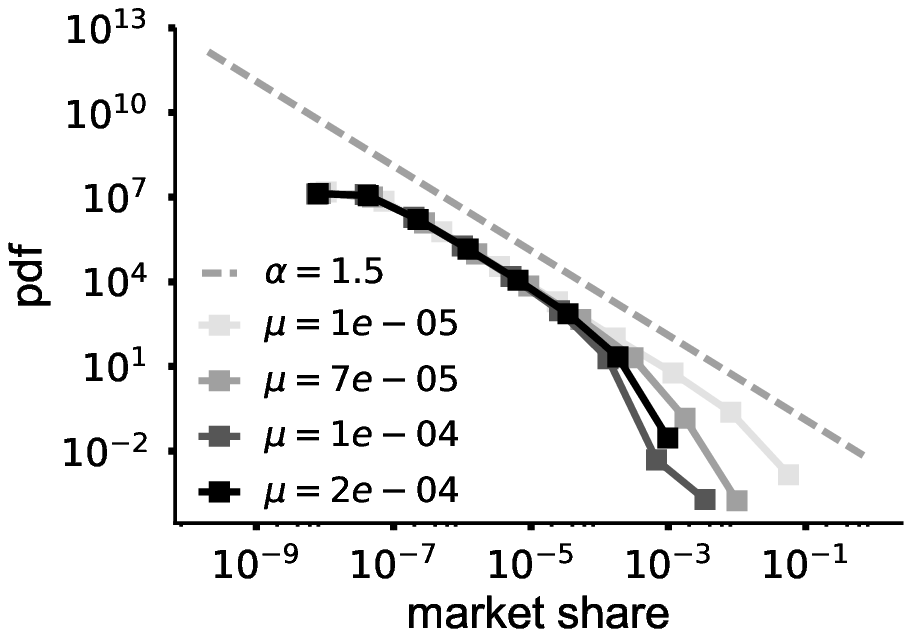}
  \end{center}
  \caption{\label{fig:DiffMue} \textbf{Distribution of species sizes for
      different values of $\mu$.}  Distribution of the species sizes
    resulted form numerical simulations given different values of $\mu$.}
\end{figure}

Our choice of the mutation parameter $\mu$ is informed by the data to yield a number of new
cryptocurrencies per unit time corresponding to the empirical
observation. By choosing $\mu = \frac{7}{N}$, where $N$ is the
population size in the model it holds that $1$ model generation
corresponds to $1$ week of observation (since on average $7$ new
cryptocurrencies enter the system every week, see Sec. \ref{sec:stability}). In Fig.~\ref{fig:DiffMue} we show that the distribution of
species sizes (see Fig. \ref{fig:MSFirstSim}A) has a very similar shape for a broad range of choices of $\mu$
\cite{O2017novelty}.

All simulations are run starting with one species in order to capture
the initial dominance of Bitcoin in the cryptocurrency market. This
reflects the initial state of the cryptocurrencies market, when Bitcoin
was the only existing cryptocurrency. Simulations are run using
$N=10^5$, implying that an individual in the model maps to
$\sim \$100,000$ (We verified that results do not depend on the choice of $N$, as long as $N$ is large enough).
 
While in the neutral model a new species enters the system as a new
individual, we further inform the model with the average size of a new
cryptocurrency ($\sim \$1.5$ million), corresponding to $m=15$ individuals in the model when $N=10^5$ as in our
case.  To consider the fact that new cryptocurrencies do not enter the
market with exactly the same size, in our simulations, when a mutation
occurs, the new species enters with a number $m$
of individuals randomly extracted from the interval $[10,20]$.

\begin{figure*}[!h]
  \begin{center}
    \includegraphics[width=5in]{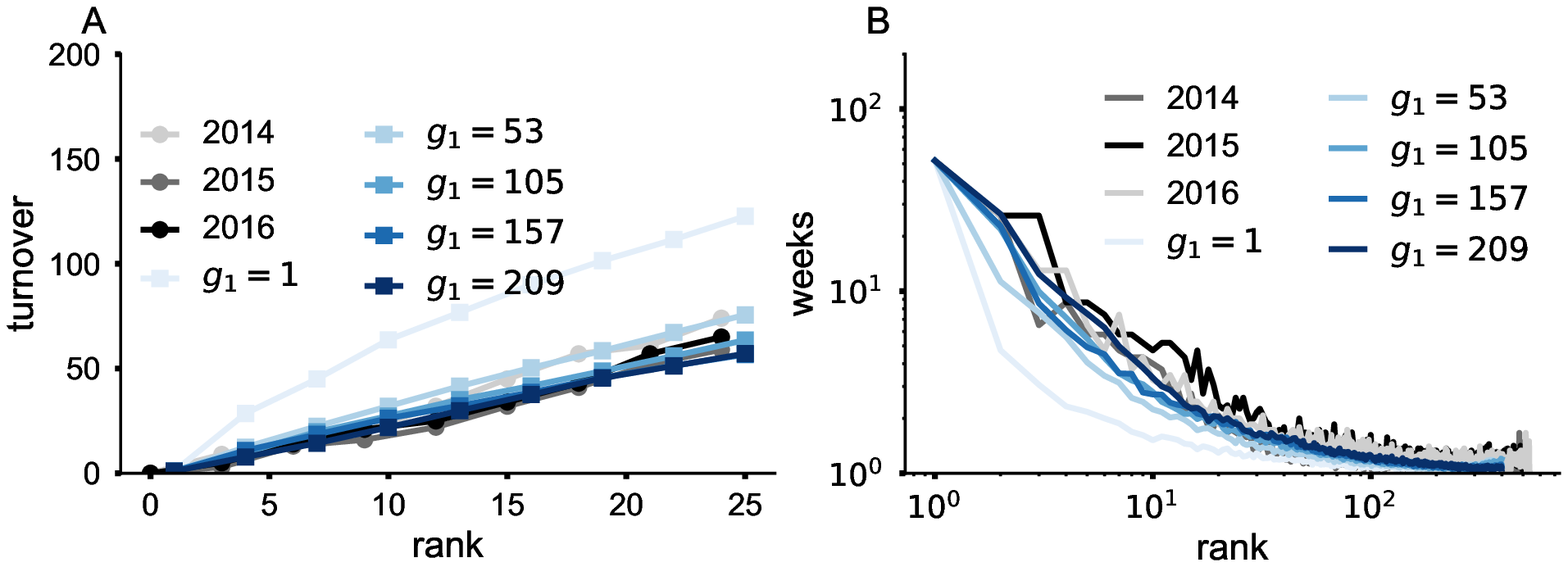}
  \end{center}
  \caption{\label{fig:TurnoverAndLifeSim} \textbf{Neutral model ranks
      dynamics.} \textbf{(A)} Turnover profile computed considering $52$
    for the cryptocurrencies data (gray lines, dots) and for numerical
    simulations (blue lines). \textbf{(B)} The Average life time a
    cryptocurrency/species stays in a given rank computed considering
    $52$ generations for the cryptocurrencies data (gray lines, dots)
    and for numerical simulations (blue lines). Simulation parameters
    are $\mu = 7/N$, $N = 10^{5}$ and 1 species in the initial state.  }
\end{figure*}

The exponent $\alpha =1.5$ exhibited by the data and the simulations(see Fig.\ref{fig:MSFirstSim}A) are equilibrium
properties of the neutral model, and hence obtained under a broad range
of conditions (e.g., initial condition, time of start of measure and
aggregation window) and robust to changes in the value of $\mu$ \cite{O2017novelty}, Fig.~\ref{fig:DiffMue}).
Fig.\ref{fig:MSFirstSim}B and C are obtained starting from generation $104$ and aggregating over $52$
generations (i.e. performing the analysis over the single population
obtained by aggregating the $N*52$ individuals
\cite{hahn2003drift,Bentley2004random}). Fig.~\ref{fig:TurnoverAndLifeSim}
shows the turnover profile (A) and average life time of a rank (B) when the
measure is performed over $52$ generations starting from different
generations $g_1$ corresponding to the first year (measures start at
generation $g_1=1$), second year (measures start at generation $g_1=53$),
etc. It is clear that, with the exception of a high rank mobility characterizing the very first generations,
the choice of $g_1$ has little effect on the curves produced by the model.
Fig.\ref{fig:MSFirstSim}D is measured from generation
$1$ up to generation $210$, corresponding to $~4$ years. Each point of
the simulation curve corresponds to the instantaneous market share of
the dominating cryptocurrency at that generation.

\subsection{technologies, same distribution}
\label{sec:PoWPoS}
In order to check whether technical differences leave any detectable
fingerprint at the level of statistical distributions, we look at
cryptocurrencies adopting one of the two main blockchain algorithms for
reaching consensus on what block represents recent transactions across
the network: Proof-of-work (PoW) or the Proof-of-stake (PoS) consensus
algorithms.

The PoW scheme was introduced as part of Bitcoin in 2009
\cite{nakamoto2008bitcoin}. To generate new blocks, participating users
work with computational and electrical resources in order to complete
``proof-of-works'', pieces of data that are difficult to produce but easy
to verify. Block generation (also called ``mining'') is rewarded with
coins. To limit the rate at which new blocks are generated, every 2016
blocks the difficulty of the computational tasks changes
\cite{franco2014understanding}.

While the PoW mechanism is relatively simple, there are concerns
regarding its security and sustainability. First, severe implications
could arise from the dominance of mining pools controlling more than
50\% of the computational resources and who could in principle
manipulate the blockchain transactions. This scenario is far from being
unrealistic: in 2014, one mining pool (Ghash.io) \cite{Ghashio}
controlled $42\%$ of the Bitcoin mining power. Also, the energy
consumption of PoW based blockchain technologies has raised
environmental concerns: it is estimated that Bitcoin consumes about
12.76 TWh per year \cite{BitcoinEnergy}.

The PoS scheme was introduced as an alternative to PoW. In this system,
mining power is not attributed based on computational resources but on
the proportion of coins held. Hence, the richer users are more likely to
generate the next block. Miners are rewarded with the transactions
fees. While proof-of-work relies heavily on energy, proof-of-stake
doesn't suffer from this issue. However, consensus is not guaranteed
since miners sole interest is to increase their profit. Through the
years both protocols have been altered to fix certain issues and
continue to be improved.

\begin{figure}[!h]
  \begin{center}
    \includegraphics[width=2.5in]{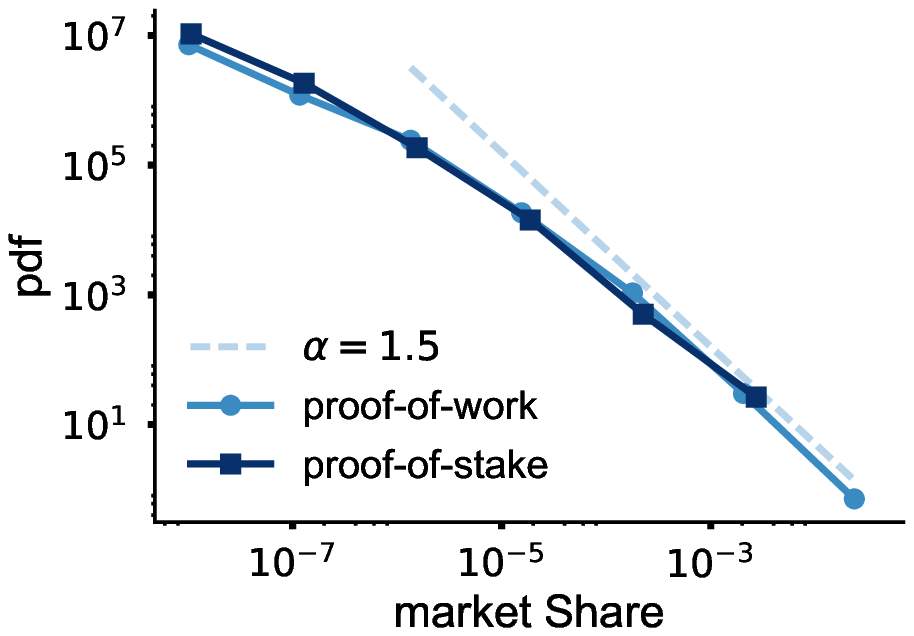}
  \end{center}
  \caption{\label{fig:PoWPoS} \textbf{Distribution of market share.}
    Distribution of the market share for proof-of-work cryptocurrencies
   (light blue filled line) and distribution of market share of (proof-of-stake or hybrid) cryptocurrencies (dark blue filled line). The dashed
    line is power law curve with exponent $\alpha = 1.5$.}
\end{figure}

Figure~\ref{fig:PoWPoS} shows that the market shares of the two groups of cryptocurrencies follow the same
behavior. The figure is generated using data collected from
\cite{MapofCoins} and \cite{coincap}.

\subsection{share and frequency-rank distributions for individual years}
\label{ver_exp}
The power-law fit for the distribution of market share (Table~\ref{tab:table2}) and the frequency-rank distribution (Table~\ref{tab:table3}) are consistent with the theoretical predictions of the neutral model\cite{adamic2002zipf} also for individual years. Fits coefficient for the distribution of market share are computed using the methodology described in \cite{clauset2009power} (errors are obtained by bootstrapping 1000 times). Fit coefficients with errors for frequency-rank distributions are computed with the least-square method. \\

\begin{table*}
\centering
\caption{\label{tab:table2}  \textbf{Power-law fit coefficients of the market share distributions.} }
\begin{tabular}{ l l } 
 \hline
\textbf{Year} & \textbf{$\alpha$}  \\
\hline
2013 & $1.37\pm 0.04 $\\
2014       &$1.54\pm 0.09 $ \\
2015       &$1.62\pm 0.12 $ \\
2016       &$1.59\pm 0.13 $ \\
2017       &$1.60\pm 0.21 $ \\
all years  &$1.58\pm 0.12 $ \\
\end{tabular}
\end{table*}
\begin{table*}
\centering
\vspace{-10cm}
\caption{\label{tab:table3}  \textbf{Power-law fit coefficients of the frequency-rank distributions.} }
\begin{tabular}{ l l } 
 \hline
\textbf{Year} & \textbf{$\beta$}  \\
\hline
2013      & $-1.98\pm 0.20 $\\
2014      & $-2.00\pm 0.13 $\\
2015      & $-1.83\pm 0.08 $\\
2016      & $-1.88\pm 0.08 $\\
2017      & $-1.86\pm 0.16 $\\
all years & $-1.93\pm 0.23 $\\
\end{tabular}
\end{table*}

\end{document}